\documentclass[12pt]{article}

\usepackage{bbold}
\usepackage{amssymb}
\usepackage{graphics}

\def\ve{\varepsilon}
\def\SS{\ss}

\def\ve{\varepsilon}

\def\r{\rho}
\def\s{\sigma}

\def\d{\partial}
\def\m{\mu}
\def\n{\nu}

\def\be{\begin{equation}}
\def\ee{\end{equation}}
\def\beq{\begin{equation}}
\def\eeq{\end{equation}}
\def\bea{\begin{eqnarray}}
\def\eea{\end{eqnarray}} 
\def\beqa{\begin{equation}\begin{array}{l}}
\def\eeqa{\end{array}\end{equation}}

\def\eqn#1{(\ref{#1})}

\def\eqref#1{eq.~(\ref{eq:#1})}

\def\nn{\nonumber}

\begin{document}

\thispagestyle{empty}
\begin{flushright}
\framebox{\small BRX-TH~548
}\\
\end{flushright}

\vspace{.8cm}
\setcounter{footnote}{0}
\begin{center}
{\Large{\bf 
Conformal Invariance of Partially Massless Higher Spins}
    }\\[10mm]

{\sc S. Deser$^\heartsuit$
and A. Waldron$^\clubsuit$
\\[6mm]}

{\em\small  
$^\heartsuit$Physics Department, Brandeis University, Waltham,
MA 02454, 
USA\\ {\tt deser@brandeis.edu}}\\[5mm]
{\em\small  
$^\clubsuit$Department of Mathematics,
University of California, Davis, CA 95616,
USA\\ {\tt wally@math.ucdavis.edu}}\\[5mm]

\bigskip

\bigskip

{\sc Abstract}\\
\end{center}

{\small
\begin{quote}

We show that there exist conformally invariant theories
for all spins in $d=4$ de Sitter space, namely the partially 
massless models with higher derivative gauge invariance under 
a scalar gauge parameter.
This extends the
catalog from the two known gauge models -- Maxwell and  partially massless
spin~2 -- to all spins.

\bigskip

\bigskip

\end{quote}
}

\newpage





\section{Introduction}

In this Letter
we present new conformally invariant gauge theories
in dimension four constant curvature spaces
generalizing the well known
conformally improved scalar and Maxwell fields. 
Conformal flatness of these spaces
then also implies lightcone propagation~\cite{Deser:1983tm}. An obvious
mechanism to achieve lightlike excitations is gauge invariance,
and indeed we recently presented new partially massless
gauge invariant higher spin theories that propagate on the lightcone
in constant curvature backgrounds. While gauge invariance alone
is certainly not sufficient to ensure conformal invariance, 
it turns out that in dimension four, there exists a series 
of gauge invariant higher spin theories whose first two elements
are the improved scalar and Maxwell fields. These theories
enjoy higher derivative ($s$ derivatives at spin $s$) gauge invariances,
with a scalar gauge parameter, just like their Maxwellian cousin. 
The ``maximal depth'' partially massless theories describe $2s$
lightlike and unitary (for $\Lambda>0$ -- de Sitter space)
physical degrees of freedom.
Unlike previous attempts~\cite{Gover} to generalize conformal invariance to 
Maxwell-like theories involving higher derivative actions,
we maintain the physical requirement of actions quadratic in derivatives
and instead increase the number of derivatives appearing in gauge variations.

The results are organized as follows: Section~\ref{HS}
contains a brief review of massive higher spins in 
constant curvature backgrounds while Section~\ref{A}
deals with action principles for these fields. Conformal
symmetry in de Sitter backgrounds is discussed in Section~\ref{conf}
and conformally improved scalars are revisited in Section~\ref{scal}.
The main result is contained in Section~\ref{cpmhs} which derives
the conformal invariance of maximal depth partially massless fields.
Lightlike propagation is displayed in Section~\ref{null}.
Our conclusions may be found in the final Section.

\section{Higher Spins in Constant Curvature Spaces}
\label{HS}

Constant curvature \be
R_{\m\n}{}^{\r\s}=-\frac{2\Lambda}{n}\
\delta^{\r}_{[\m}\delta^{\s}_{\n]}\, 
\ee
spaces in $d\equiv n+1$ dimensions
are a consistent background for higher spin propagation. The
cosmological constant $\Lambda$ is positive for de Sitter space in our
conventions. We will often work in the steady state patch
\be
ds^2=-dt^2+e^{2Mt} d\vec x^{\, 2}\, ,\label{steady}
\ee
where
\be
M^2\equiv\frac{\Lambda}{n}\, .
\ee

The field equation for physical fields is~\cite{Deser:2001xr}
\be
\Big(D_\mu D^\mu +\Big[(s-1)^2+(s-2)(n-3)-3\Big] M^2-m^2\Big)\, 
\varphi^{\rm phys}_{\mu_1\ldots\mu_s}=0\, .
\label{eom}
\ee
Physical fields are classified by the mass parameter $m$. In steady
state coordinates they obey the following criteria for all $m^2>0$
\be
\varphi_{0\mu_2\ldots\mu_s}=0=\varphi^\rho{}_{\rho\mu_2\ldots\mu_s}\,
,\label{temporal}
\ee
as well as additional conditions 
\be
\d^{\mu_1}\ldots\d^{\mu_t}\varphi_{\mu_1\ldots\mu_s}=0\, , 
\label{helicities}
\ee
when 
\be
m^2=(t-1)(2s-t+n-3)M^2\, ,\qquad
t=1,\ldots,s\, .\label{tunes}
\ee
The integer $t$ is called the ``depth'' of a partially massless
field subject to a mass tuning~\eqn{tunes}.
Depth $t=1$ corresponds to the on-shell condition for a strictly
massless spin~$s$ field whose physical fields are spatially
traceless-transverse symmetric tensors. 
For all values of the depth $1\leq t\leq s$, the theory enjoys a 
higher derivative gauge invariance
\be
\delta \phi_{\mu_1\ldots\mu_s}=
D_{(\mu_1}\cdots D_{\mu_t}\xi_{\mu_{t+1}\ldots\mu_s)}+\cdots\, .
\ee
Non-tuned values of the mass
\be
m^2<(s-1)(s+n-3)M^2\, ,\label{mass}
\ee
correspond to massive fields described by spatially traceless tensors.
(Unitarity is violated for mass values not obeying~\eqn{tunes}
or~\eqn{mass}~\cite{Deser:2001pe}.) 
Strictly massless and massive fields survive in the
Minkowski limit $M=0$. Peculiar to de Sitter space are physical fields
for values of $t\neq1$~\cite{Deser:2001pe}. 
These are partially massless fields of depth
$t$. Like their strictly massless counterparts they propagate at speed 
of light. Their physical degrees of freedom are spanned by
intermediate helicity counts as demanded by~\eqn{helicities}.
These theories are also unitary in de Sitter ($\Lambda>0$)
backgrounds and obey an energy positivity condition within the
horizon~\cite{Deser:2001wx}.

\section{Actions}
\label{A}

Defining
\be
\varphi^{\rm phys}_{i_1\ldots i_s}\equiv\sqrt{-g}^{\
\frac{s}{n}-\frac12}q_{\ve}\, ,
\label{fieldredef}
\ee
where $\ve$ stands for a helicity 
labeling\footnote{See~\cite{Deser:2001wx} for a 
details of this Hamiltonian analysis.}  of the spatial 
indices $i_1\ldots i_s$
subject to~\eqn{temporal} and possibly~\eqn{helicities},
the field equation~\eqn{eom} becomes
\be
\Big(-\frac{d^2}{dt^2}+\Delta+
\Big[\frac{M(2s+n-4)}{2}\Big]^2-m^2\Big)q_\ve=0\, .
\label{heleqn}
\ee
Here the spatial Laplacian 
$\Delta\equiv e^{2Mt}\vec\d^2$ carries
the only explicit time dependence and we will raise and lower spatial
indices with impunity. An action principle follows
immediately
\be
S=\int dt \Big(\sum_\ve p_{\ve}\dot q_{\ve}-H\Big)\, ,
\label{action}
\ee
where the Hamiltonian
\be
H\equiv\sum_\ve 
\int d^nx\, \frac12\Big[p^2_\ve+e^{-2Mt}[\vec \d q_\ve]^2+\mu^2q_\ve^2\Big]\, . 
\ee
Here the ``effective mass'' is
\be
\mu^2\equiv m^2-[\frac{M(2s+n-4)}{2}\Big]^2\, .
\ee
The action~\eqn{action} may also be obtained by a detailed constraint
analysis of covariant, higher spin actions~\cite{Deser:2001wx}.

\section{Conformal Symmetry}
\label{conf}

De Sitter spacetime, being conformally flat, enjoys not only
an $so(d,1)$ isometry algebra but also an $so(d,2)$ conformal algebra.
Perhaps the simplest construction of explicit conformal Killing vectors 
is to express the metric in a manifestly conformally flat frame
\be
ds^2=\frac{-d\tau^2+d\vec x^2}{M^2\tau^2}\, , 
\ee
where $\tau=-M^{-1} \exp\Big(Mt\Big)$ is the conformal time coordinate.
One then writes down the flat conformal generators in the coordinates
$(\tau,\vec x)$,
\bea
i P_i=\partial_i\, ,&&i L=  \frac{d}{d\tau}\, ,\nn\\
i D= \tau \frac{d}{d\tau}+\vec x\cdot\vec\partial\, ,\qquad
i M_{ij}&=& x_i \d_j-x_j \d_i\, ,\qquad
i N_i =x_i \frac{d}{d\tau}+\tau\partial_i\, ,\nn\\ 
\nn\\
iK_i=2i x_i D+\Big[-\tau^2 + \vec x^2\Big]\ \partial_i\, , \!\!\!&&\!\!\!
i J =-2i\tau D+\Big[-\tau^2 + \vec x^2\Big]\ \frac{d}{d\tau}\, .\nn\\
\eea
The generators $(P_i,D,M_{ij},K_i)$ obey the $so(d,1)$ de Sitter isometry
algebra while the remaining generators enlarge this to the conformal
$so(d,2)$ algebra with corresponding conformal Killing vectors
\be
[iL,ds^2]=-\frac{2}{\tau}\ ds^2\, ,\quad
[iN_i,ds^2]=-\frac{2x_i}{\tau}\ ds^2\, ,\quad
[iJ,ds^2]=-\frac{2(-\tau^2+\vec x^2)}{\tau}\ ds^2\, .
\ee
In general we will denote the function  multiplying $2ds^2$ 
on the right hand side (equaling the divergence of
the corresponding conformal Killing vector divided by the
dimensionality of spacetime) as $\alpha_X$:
\be
[iX,ds^2]=2\alpha_X\ ds^2\, .
\ee

It is important to  note that the conformal $so(d,2)$ algebra 
is obtained by requiring closure of the generator $L$ and the
$so(d-1,1)$ isometry
algebra under commutation. This has the pleasant consequence
that $L$ and de Sitter invariance are sufficient  for a theory 
to be conformal. In the steady state coordinates~\eqn{steady}
we have
\be
iL=e^{Mt}\frac{d}{dt}\, .
\ee

\section{Scalars}
\label{scal}

The improved scalar action
\be
S=-\frac12\int dx \sqrt{-g}\left(
\d_\m\varphi  g^{\m\n} \d_\n \varphi + \frac16 R\  \varphi^2\right)\, ,
\ee
is invariant under conformal transformations
\be
\delta\varphi = X \varphi + \Big(\frac d2-1\Big)\ \alpha_X \varphi\, ,
\label{transf}
\ee
where $X\equiv \xi^\rho \d_\rho$, $D_{(\mu}\xi_{\nu)}=\alpha_X\ g_{\m\n}=
\frac 1d D.\xi \ g_{\m\n}$ and $\frac d2-1$ is the conformal
weight of the field $\varphi$.

It is useful for our purposes to spell out this invariance explicitly 
in a de Sitter background. In the steady state coordinates, the
action reads
\be
S=-\frac{1}{2}\int dt d^nx\ e^{nMt}\left(-\dot\varphi^2+
e^{-2Mt}[\vec\d\varphi]^2+\frac{M^2}{4}\ (n^2-1)\varphi^2\right)\, .
\label{dSscalar}
\ee
As discussed in the previous Section, it suffices
to consider the generator $iL=e^{Mt}\frac{d}{dt}$.
By either computing the right hand side of~\eqn{transf}
or examining the gradient terms in the action, which must be
separately conformally invariant, we find
\be
\delta_L \varphi=e^{(3-n)Mt/2} \frac{d}{dt}\Big(
e^{(n-1)Mt/2}\varphi\Big)\, .
\ee
It is then easy to verify that the $\delta_L$ variation of the Lagrangian
in~\eqn{dSscalar} is a total time derivative.

Let us now perform this computation yet again in a first
order formulation. Making the field redefinition~\eqn{fieldredef}
and a Legendre transformation, the action reads
\be
S=\int dt dx^n\left(
p\dot q-\frac12\left[
p^2+e^{-2Mt}[\vec\d q]^2-\frac14M^2q^2
\right]
\right)\, .
\label{the1}
\ee
It enjoys the conformal invariance
\be
\delta_L q=e^{3Mt/2}\frac{d}{dt}\Big(
e^{-Mt/2}q\Big)\, ,\qquad
\delta_L p=\frac{d}{dt}\Big(e^{3Mt/2}\frac{d}{dt}\Big(
e^{-Mt/2}q\Big)\Big)\, .
\ee
Although the above calculation is a triviality, the conformally
invariant action~\eqn{the1} plays an archetypal r\^ole in what follows.

\section{Conformal Partially Massless Higher Spins}
\label{cpmhs} 

Our remaining task is to determine which, if any, of the
partially massless theories are also conformal. A Herculean, 
but perhaps noble task would be to write down covariant
actions at arbitrary values of the spin\footnote{Indeed, covariant
actions can be determined from the work of~\cite{Zinoviev:2001dt}.} 
and determine 
conformal invariance explicitly. To arrive quickly at an answer,
however, we proceed as follows. Firstly, we need only look
for invariance with respect to ``de Sitter dilations'' $\delta_L$.
Secondly, a notable feature of free higher spin fields is
that despite the complexity of their covariant actions,
first order action principles in terms of physical degrees
of freedom are extremely simple--see equation~\eqn{action}!

Now, the crux of our argument: compare actions~\eqn{action} and~\eqn{the1}.
Since the dilation $\delta_L$ necessarily has no spin dependence, 
we can ignore the helicity summation $\sum_\ve$. Therefore conformal 
invariance is guaranteed by choosing the effective mass 
\be
\mu^2=-\frac{1}{4}M^2\, .
\ee
Setting the mass parameter $m^2$ to its depth $t$ partially massless
value we therefore require
\be
-\frac{1}{4}M^2(2s-2t+n-2)^2=-\frac{1}{4}M^2\, ,
\ee
which is solved via
\be
t=s+\frac{n\pm 1}{2}-1\, .
\ee
However, the depth $t$ must be both integer and no greater than $s$.
The only the solutions therefore are $n=3,1$, {\it i.e.} spacetime
dimensions four and two!
In dimension two, there are no helicities so we find nothing new.

Dimension four is more interesting, we are forced to take $t=s$.
When $s=t=0$, we obtain a conformally improved scalar. For $s=t=1$,
we have a single gauge invariance
\be
\delta\phi_\mu=D_\mu \xi\, ,
\ee
yielding Maxwell theory in four dimensional 
de Sitter space with its scalar gauge parameter.
This theory is well known
to be conformal. For $s=t=2$, we have a double derivative gauge invariance
\be
\delta\phi_{\mu\nu}=\Big(D_{(\mu}D_{\nu)}+\frac\Lambda3\Big)\ \xi\, 
\ee
and we obtain the original spin~2 partially massless theory 
of~\cite{Deser:1983tm}. Indeed, these authors arrived
at this theory by demanding conformal invariance and found a higher
derivative gauge transformation as a consequence, as opposed to 
the opposite logic which led to partially massless yet not necessarily
conformal theories 
in~\cite{Deser:2001xr,Deser:2001pe,Deser:2001wx,Deser:2003gw}.

In general, our result is that in dimension four, partially massless
theories with maximal depth gauge invariances
\be
\delta \phi_{\mu_1\ldots\mu_s}=\Big(D_{(\mu_1}\ldots D_{\mu_s)}+\cdots\Big)\
\xi\, ,
\ee
are conformally invariant.

\section{Null Propagation} 
\label{null}

Another way to uncover the conformal invariance of 
dimension four maximal depth partially massless theories
is to solve explicitly their higher spin wave equations.
This computation has been carried out in~\cite{Deser:2001xr}.
In steady state coordinates the wave equations for these theories
are of Bessel type. In general the solutions are, of course,
Bessel functions, but Huygen's principle~\cite{Helgason} implies that 
for half integer values of their index $\nu$, these Bessel solutions
become simply massless plane waves multiplied by slowly varying
polynomials. There are cardinally infinitely many such special solutions
but the index $\nu=1/2$ is special, since it corresponds to a
conformally improved scalar. Some details:

Let us work in four dimensional spacetime since that is 
where the interesting conformal theories
live. Fourier transforming $\vec \d\rightarrow i\vec k$ and 
rescaling the conformal time coordinate $\tau\rightarrow z=|\vec
k|\tau$, 
the field equation for a massive spin $s$, helicity $\ve$ field in de 
Sitter space reads
\be
\frac{d^2q}{dz^2}+\frac1z+\Big(1-\frac{\nu^2}
{z^2}\Big)q=0,
\ee
where $q\equiv |\vec k/M|^{3/2-s}q_\ve$ and the index
\be
\nu^2=\frac14+s(s-1)-\frac{m^2}{M^2}\, .
\ee
Setting $m^2$ to its tuned values~\eqn{tunes} yields
\be
\nu^2=\frac14(2s-2t+1)^2\, .
\ee
The conformal value $\nu^2=1/4$ is obtained only for $s=t$ in agreement with
the analysis of the previous Section. In this case the solution to the wave 
equation is $q(z)=z^{-1/2}\exp(iz)$ which amounts to massless
plane wave propagation since the overall exponential behavior
is $\exp(i|\vec k|\tau+i\vec k\cdot\vec x)$.

\section{Conclusions}

In this Letter we have uncovered new conformal, gauge invariant
theories in four dimensional de Sitter space 
generalizing Maxwell's vector theory. These new theories describe
spin $s$, lightlike excitations with helicities $\pm s,\ldots,\pm 1$.
As for Maxwell theory,  elimination of the zero 
helicity mode via gauge invariance suffices to ensure conformal invariance.

Many pressing questions remain but as usual the main difficulty is
interactions. As yet no obvious mechanism for partially massless
interactions is available, although one might expect
Strings to do the job~\cite{Dolan:2001ih}. Another speculation is 
that since the maximal depth partially massless theories
are singled out by conformal invariance, perhaps the interaction
problem is more tractable for these theories. The guiding principle
would, of course, be conformal invariance.

\section*{Acknowledgments}
A.W. thanks the Max Planck Institut f\"ur Mathematik Bonn
for hospitality and the Vivatsgasse {\it Stra{\SS}enmusiker}
for inspiration.
This work was supported by the 
National Science Foundation under grants PHY04-01667
and PHY-01-40365.

\end{document}